\documentclass[twocolumn,aps,superscriptaddress,showpacs,a4paper]{revtex4}
\usepackage{graphicx} 
\usepackage{amsmath,amssymb}
\usepackage{dcolumn}

\newcommand{\force}{\mathcal{A}}
\newcommand{\flux}{\mathcal{J}}
\newcommand{\bN}{\bar{N}}
\newcommand{\brho}{\bar{\rho}}

\begin{document} 
\title{Fluctuation theorem for entropy production during effusion of a
  relativistic ideal gas}  
\author{B. Cleuren}
\affiliation{Hasselt University - B-3590 Diepenbeek, Belgium}
\author{K. Willaert}
\affiliation{Hasselt University - B-3590 Diepenbeek, Belgium}
\author{A. Engel}
\affiliation{Universit\"{a}t Oldenburg, Institut f\"{u}r Physik, 26111
  Oldenburg, Germany}
\author{C. Van den Broeck}
\affiliation{Hasselt University - B-3590 Diepenbeek, Belgium}

\begin{abstract}
The probability distribution of the entropy production for the effusion of a relativistic ideal gas is calculated explicitly. This result is then extended to include particle and anti-particle pair production and annihilation. In both cases, the fluctuation theorem is verified.   
\end{abstract}
\date{\today}
\maketitle

\section{introduction}
The fluctuation theorem \cite{evans1,gallavotti,gallavotti2,comren}
can be seen as a natural extension to far from equilibrium situations
of the pioneering work of Onsager on the reciprocity relations. The
theorem has been proven in a variety of settings, including
thermostated systems \cite{evans3,dolo,gilbert}, Markov processes
\cite{kurchan,lebowitz,maes,gaspard1,gaspard2}, classical Hamiltonian 
dynamics \cite{jarzynski,cleurenPRL}, quantum dynamics, quantum field theory
\cite{esposito1,esposito2,esposito3} and relativity \cite{fingerle}. The
wide range of validity of the theorem reflects the fact that 
it derives from two basic ingredients of the fundamental laws of nature: time
reversibility and time translational invariance. The  effusion of gases
provides a simple setting in which the fluctuation theorem can be verified on
the basis of freshmen's physics \cite{cleurenPRE2006,woodPRE2007}. The
effusion of radiation (photons) can be discussed on a similar basis
\cite{cleurenEPL2007}. In this paper, we show that the fluctuation
theorem is also valid for the effusion of classical relativistic particles, including particle and anti-particle pair production and annihilation.

\section{fluctuation theorem for effusion of a relativistic ideal gas}
Two relativistic ideal gases, $A$ and $B$, are initially in equilibrium at
given densities $\rho_{i}$ and temperature $T_{i}$, $i \in\{A,B\}$
while separated by a common adiabatic wall. At time $t=0$,
a small hole with surface area $\sigma$ is opened, allowing an
exchange of particles and energy between the two gases during a
(fixed) time interval of length $t$. We assume the presence of an
energy filter, so that only particles with total energy in the range $[
E_{1},E_{2}]$ can cross.\newline
The dimensions of the hole are supposed to be small compared to the mean free path of the gas particles, and
the reservoirs sufficiently large so that the equilibrium state in
both parts is not affected by the exchange. In particular, the
temperatures and densities in both reservoirs remain constant. Hence,
the change in entropy in the total system, upon a total transfer of
energy $\Delta U$ and of particles $\Delta N$ from $A$ to $B$ during
$t$, is given by:  
\begin{eqnarray}
\Delta S&=& \Delta S_{A}+ \Delta S_{B}
\nonumber \\&=&-\frac{1}{T_{A}}\Delta U+\frac{\mu_{A}}{T_{A}}\Delta
N+\frac{1}{T_{B}}\Delta U-\frac{\mu_{B}}{T_{B}}\Delta N \nonumber
\\&=&\force_{U}\Delta U +\force_{N}\Delta N. 
\label{eq:entropy}
\end{eqnarray}
We introduced, in accordance with the definitions from irreversible
thermodynamics, the following thermodynamic forces (affinities) for
energy and particle flow, respectively: 
\begin{equation}\label{eq:forces}
\force_{U}=\frac{1}{T_{B}}-\frac{1}{T_{A}} \;\;\;\; ; \;\;\;\;
\force_{N}=\frac{\mu_{A}}{T_{A}}-\frac{\mu_{B}}{T_{B}}. 
\end{equation} 
A derivation of the chemical potential $\mu_{i}$ of a relativistic
ideal gas is given in Appendix \ref{thermo}, and leads to the
following explicit expression for $\force_{N}$: 
\begin{equation}
\force_{N}=k\log\left[\frac{\rho_{A}T_{B}K_{2}(mc^{2}/kT_{B})}{\rho_{B}T_{A}K_{2}(mc^{2}/kT_{A})}\right],
\label{eq:tforces}
\end{equation}
where $K_{2}$ is the (second) modified Bessel function of the second kind.
In writing down these results, we use the convention that the internal
energy $U$ of the gas comprises both the kinetic energy and the rest
energy of the particles $(mc^{2})$. Note that the thermodynamic force
$\force_{N}$ diverges when one of the (infinitely large!) reservoirs
is empty, so that free effusion into unlimited space, implying an
infinitely large entropy production, corresponds to a singular limit.

We will show below that the joint probability density $P_{t}(\Delta
U,\Delta N)$ satisfies the following detailed fluctuation theorem: 
\begin{equation}
\frac{P_{t}(\Delta U,\Delta N)}{P_{t}(-\Delta U,-\Delta N)}=e^{\Delta S/k}.
\end{equation}
with $\Delta S$ given by Eq.~(\ref{eq:entropy}). This encompasses as a special case the more familiar form of the fluctuation theorem:
\begin{equation}
\frac{P_{t}(\Delta S)}{P_{t}(-\Delta S)}=e^{\Delta S/k}.
\end{equation}
Since the increments of $\Delta U$ and $\Delta N$ in successive time intervals
are independent, one has that:  
\begin{multline}
\sum_{\Delta N}\int e^{-(\lambda_{U}\Delta U+\lambda_{N}\Delta
  N)}P_{t}(\Delta U, \Delta N)d\Delta U=\\ e^{-tQ(\lambda_{U},\lambda_{N})}. 
\end{multline}
where $Q(\lambda_{U},\lambda_{N})$ is the cumulant generating function. In terms of the latter, the detailed fluctuation theorem reads: 
\begin{equation}
Q(\lambda_{U},\lambda_{N})=Q(\force_{U}/k-\lambda_{U},\force_{N}/k-\lambda_{N}).
\label{DFT}
\end{equation}

\section{Master equation and cumulant generating function}
During a small time interval $dt$, the contributions to quantities $\Delta U$ and $\Delta N$ result from individual particle transport across the hole. Following basic arguments from kinetic theory of gases the corresponding probabilities per unit time, $T_{A \rightarrow B}(E)$ and $T_{B \rightarrow A}(E)$, to observe a particle with energy $E$ crossing the hole from $i \rightarrow j$ is given by (cf. Appendix \ref{transition}):
\begin{equation}\label{eq:tr}
T_{i \rightarrow j}(E)= \frac{\sigma \rho_{i}}{4c^{2}(mc)^{3}}\frac{(mc^{2}/kT_{i})e^{-\frac{E}{kT_{i}}}}{K_{2}(mc^{2}/kT_{i})}(E^{2}-m^{2}c^{4}).
\end{equation} 
The probability density $P_{t}(\Delta U,\Delta N)$ thus obeys the following Master equation:
\begin{multline}
\frac{\partial}{\partial t}P_{t}(\Delta U,\Delta N)=\\ \int_{E_{1}}^{E_{2}} T_{A\rightarrow B}(E)P_{t}(\Delta U-E,\Delta N-1)dE \\+\int_{E_{1}}^{E_{2}} T_{B\rightarrow A}(E)P_{t}(\Delta U+E,\Delta N+1)dE  \\
-P_{t}(\Delta U,\Delta N)\int_{E_{1}}^{E_{2}}\left[T_{A \rightarrow B}(E)+T_{B \rightarrow A}(E)\right]dE,
\label{eq:master}
\end{multline}
subject to the initial condition $P_{0}(\Delta U,\Delta N)=\delta (\Delta U)\delta_{\Delta N,0}$.\newline The cumulant generating function $Q(\lambda_{U},\lambda_{N})$ is obtained by multiplying the master equation with $\exp\{-\lambda_{U}\Delta U-\lambda_{N}\Delta N\}$ and subsequent integration over $\Delta U$ and summation over $\Delta N$ respectively. This leads to the following analytical result:
\begin{multline}
Q(\lambda_{U},\lambda_{N})=I_{A}(0,0)-I_{A}(\lambda_{U},\lambda_{N})\\+I_{B}(0,0)-I_{B}(-\lambda_{U},-\lambda_{N}),
\label{eq:mu}
\end{multline}
where we define:
\begin{equation}
I_{i}(\lambda_{U},\lambda_{N})=e^{-\lambda_{N}}\int_{E_{1}}^{E_{2}} T_{i \rightarrow j}(E)e^{-\lambda_{U}E}dE. 
\end{equation}
The evaluation of the integral yields:
\begin{multline}\label{eq:resI}
I_{i}(\lambda_{U},\lambda_{N})=\frac{e^{-\lambda_{N}}\sigma \rho_{i}}{4c^{2}(mc)^{3}}\frac{mc^2/kT_{i}}{K_{2}(mc^2/kT_{i})}\left(\frac{kT_{i}}{1+\lambda_{U}kT_{i}}\right)^{3} \\ \times\Bigg[
e^{-\left(\frac{1}{kT_{i}}+\lambda_{U}\right)E_{1}}\bigg\{2+2E_{1}\left(\frac{1}{kT_{i}}+\lambda_{U}\right)\\ +(E_{1}^{2}-m^{2}c^{4})\left(\frac{1}{kT_{i}}+\lambda_{U}\right)^{2}\bigg\} 
\\-e^{-\left(\frac{1}{kT_{i}}+\lambda_{U}\right)E_{2}}\bigg\{2+2E_{2}\left(\frac{1}{kT_{i}}+\lambda_{U}\right)\\ +(E_{2}^{2}-m^{2}c^{4})\left(\frac{1}{kT_{i}}+\lambda_{U}\right)^{2}\bigg\}\Bigg].
\end{multline}
Eq.~(\ref{eq:mu}), together with Eq.~(\ref{eq:resI}), is the central result of this
paper. A tedious calculation, using the explicit expressions for the
thermodynamic forces given in Eq.~(\ref{eq:tforces}), reveals that
this expression indeed verifies the detailed fluctuation theorem
Eq.~(\ref{DFT}).\newline
The analysis becomes somewhat simpler in the extreme relativistic limit $mc^2/kT\ll 1$. In this case the relativistic energy becomes $H(p)=cp$ and the results from Appendices \ref{thermo} and \ref{transition} simplify:
\begin{eqnarray}
\label{eq:Zrel}
Z(T,V,N)&=&\frac{1}{N!}\left[\frac{8\pi V}{(hc/kT)^3}\right]^N;\\
\label{eq:Frel}
F(T,V,N)&=&-NkT\left[\ln\frac{8\pi}{(hc/kT)^3}\frac{V}{N} +1 \right];\\
\label{eq:trrel}
T_{i \rightarrow j}(E)&=& \frac{\sigma \rho_{i}c}{8(kT_i)^{3} }E^2
e^{-\frac{E}{kT_{i}}};\\
\force_{N}&=&k\log\left[\frac{\rho_{A}T_{B}^3}{\rho_{B}T_{A}^3}\right].
\label{eq:tforcesrel}
\end{eqnarray}
The same expressions result of course also from (\ref{eq:Z}), (\ref{eq:F}),
(\ref{eq:tr}) and (\ref{eq:tforces}) respectively by
using $E_i\gg mc^2$ and $K_2(u)\sim 2/u^2$ for $u\ll 1$. Eq.~(\ref{eq:resI}) simplifies to
\begin{equation}
\label{eq:resIrel}
I_i(\lambda_U,\lambda_N)=\frac{ \sigma\rho_i c}{4(1+\lambda_U k T_i)^3} e^{-\lambda_N}\; .
\end{equation}
and the cumulant generating function becomes
\begin{multline}
  \label{eq:murel}
  Q(\lambda_U,\lambda_N)=\frac{\sigma\rho_1 c}{4}
   \Big[1-\frac{e^{-\lambda_N}}{(1+\lambda_U k T_1)^3}\Big]\\
  +\frac{\sigma\rho_2 c}{4}
   \Big[1-\frac{e^{-\lambda_N}}{(1+\lambda_U k T_2)^3}\Big].
\end{multline}
The fluctuation theorem (\ref{DFT}) is easily verified.  
\section{Cumulants}
A Taylor expansion of the cumulant generating function gives
immediately the expressions of the joint cumulants $\kappa_{ij}$ (of
power $i$ in the energy flux and power $j$ in the particle flux): 
\begin{equation}
Q(\lambda_{U},\lambda_{N})=-\frac{1}{t}\sum_{i,j}\frac{(-1)^{i+j}\lambda_{U}^{i}\lambda_{N}^{j}}{i!j!}\kappa_{ij}.
\end{equation}
As the higher order expressions quickly become unwieldy, we will only present the first order results, namely $\kappa_{10}=\langle \Delta U\rangle$ and $\kappa_{01}=\langle \Delta N\rangle$. These results will be used in the next section to validate the Onsager symmetry.
\begin{widetext}
\begin{multline}\label{eq:k10}
\langle \Delta U\rangle= \frac{t \sigma \rho_{A} mc^{2}}{4m^{3}c^{5}K_{2}(mc^{2}/kT_{A})}\bigg[e^{-\frac{E_{1}}{kT_{A}}}\Big(E_{1}^{3}-E_{1}m^{2}c^{4}+3E_{1}^{2}kT_{A}-m^{2}c^{4}kT_{A}+6E_{1}(kT_{A})^{2}+6(kT_{A})^{3}\Big) \\ -e^{-\frac{E_{2}}{kT_{A}}}\Big(E_{2}^{3}-E_{2}m^{2}c^{4}+3E_{2}^{2}kT_{A}-m^{2}c^{4}kT_{A}+6E_{2}(kT_{A})^{2}+6(kT_{A})^{3}\Big)\bigg] \\
-\frac{t \sigma \rho_{B} mc^{2}}{4m^{3}c^{5}K_{2}(mc^{2}/kT_{B})}\bigg[e^{-\frac{E_{1}}{kT_{B}}}\Big(E_{1}^{3}-E_{1}m^{2}c^{4}+3E_{1}^{2}kT_{B}-m^{2}c^{4}kT_{B}+6E_{1}(kT_{B})^{2}+6(kT_{B})^{3}\Big) \\ -e^{-\frac{E_{2}}{kT_{B}}}\Big(E_{2}^{3}-E_{2}m^{2}c^{4}+3E_{2}^{2}kT_{B}-m^{2}c^{4}kT_{B}+6E_{2}(kT_{B})^{2}+6(kT_{B})^{3}\Big)\bigg],
\end{multline}
and
\begin{multline}\label{eq:k01}
\langle \Delta N\rangle= \frac{t \sigma \rho_{A} mc^{2}}{4m^{3}c^{5}K_{2}(mc^{2}/kT_{A})}\bigg[e^{-\frac{E_{1}}{kT_{A}}}\Big(E_{1}^{2}-m^{2}c^{4}+2E_{1}kT_{A}+2(kT_{A})^{2}\Big)  -e^{-\frac{E_{2}}{kT_{A}}}\Big(E_{2}^{2}-m^{2}c^{4}+2E_{2}kT_{A}+2(kT_{A})^{2}\Big)\bigg] \\
-\frac{t \sigma \rho_{B} mc^{2}}{4m^{3}c^{5}K_{2}(mc^{2}/kT_{B})}\bigg[e^{-\frac{E_{1}}{kT_{B}}}\Big(E_{1}^{2}-m^{2}c^{4}+2E_{1}kT_{B}+2(kT_{B})^{2}\Big)-e^{-\frac{E_{2}}{kT_{B}}}\Big(E_{2}^{2}-m^{2}c^{4}+2E_{2}kT_{B}+2(kT_{B})^{2}\Big)\bigg].
\end{multline}
\end{widetext}

\section{Onsager Relation} 
The average entropy production during the effusion process is obtained by averaging Eq.~(\ref{eq:entropy}) and taking the time derivative:
\begin{equation}
\frac{d}{dt}\langle \Delta S \rangle = \force_{U}\flux_{U} +\force_{N}\flux_{N},
\end{equation}
where we introduce the macroscopic fluxes $\flux_{U}$ and $\flux_{N}$ corresponding to energy and particle transport:
\begin{equation}
\flux_{U}=\frac{d}{dt}\langle \Delta U \rangle \;\;\;\; ; \;\;\;\; \flux_{N}=\frac{d}{dt}\langle \Delta N \rangle.
\end{equation} 
These fluxes in general depend nonlinearly on the thermodynamic forces. Close to equilibrium, when the thermodynamic forces are small (that is, for small temperature and density difference), the expressions can be linearised, and written in compact matrix notation
\begin{equation}
\bar{\flux}=L\bar{\force},
\end{equation}
with $\bar{\flux}=(\flux_{U},\flux_{N})^{T}$, $\bar{\force}=(\force_{U},\force_{N})^{T}$ and $L$ the Onsager matrix :
\begin{equation}
        L=
        \begin{bmatrix}
        L_{UU} & L_{UN} \\ L_{NU} & L_{NN}      
        \end{bmatrix}.
        \label{matrix}
\end{equation}
The following explicit expression for these Onsager coefficients are obtained from Eqs.~(\ref{eq:k10}) and (\ref{eq:k01}):
\begin{widetext}
\begin{eqnarray}
L_{UU}&=& \frac{\sigma \rho}{4km^{2}c^{3}K_2\left(\frac{mc^2}{kT}\right)}\bigg(e^{-\frac{E_1}{kT}}\big({E_1}^4+4{E_1}^3kT+12{E_1}^2k^2T^2-{E_1}^2m^2c^4
 +24E_1k^3T^3 -2E_1kTm^2c^4 \nonumber \\ && \qquad\qquad\qquad\qquad\quad+24k^4T^4  -2k^2T^2m^2c^4 \big)-e^{-\frac{E_2}{kT}}\big({E_2}^4+4{E_2}^3kT +12{E_2}^2k^2T^2-{E_2}^2m^2c^4 \nonumber \\ && \qquad\qquad\qquad\qquad\qquad\qquad\qquad\qquad\qquad\qquad\qquad+24E_2k^3T^3 -2E_2kTm^2c^4 +24k^4T^4 -2k^2T^2m^2c^4 \big)\bigg)\nonumber \\
L_{UN}&=&L_{NU}=\frac{\sigma \rho}{4km^{2}c^{3}K_2\left(\frac{mc^2}{kT}\right)} \bigg(e^{-\frac{E_1}{kT}}\big({E_1}^3+3{E_1}^2kT+6E_1k^2T^2 -E_1m^2c^4 +6k^3T^3-kTm^2c^4\big)  \nonumber \\
        & & \qquad\qquad\qquad\qquad\qquad\qquad\qquad\qquad- e^{-\frac{E_2}{kT}}\big({E_2}^3+3{E_2}^2kT+6E_2k^2T^2 -E_2m^2c^4 +6k^3T^3 -kTm^2c^4\big)\bigg)\nonumber\\
L_{NN}&=&\frac{\sigma \rho}{4km^{2}c^{3}K_2\left(\frac{mc^2}{kT}\right)}\left(e^{-\frac{E_1}{kT}}\left({E_1}^2+2E_1kT+2k^2T^2-mc^2\right)-e^{-\frac{E_2}{kT}}\left({E_2}^2+2E_2kT+2k^2T^2-mc^2\right)\right).
  \label{nrel}
\end{eqnarray}
\end{widetext}
The symmetry property of the matrix is a direct consequence of the symmetry relation Eq.~(\ref{DFT}) of the cumulant generating function.\newline
These expressions significantly simplify when considering an infinitesimal energy window ($dE \ll E_{0}$):
\begin{eqnarray}
E_{1}&=&E_{0}-dE/2 \;\; ;\nonumber \\
E_{2}&=&E_{0}+dE/2 \;\; .
\end{eqnarray}
In this case, there is a strong coupling between the particle and energy current: every particle that crosses carries a energy $E_{0}$ and so $\flux_{U}=E_{0}\flux_{N}$. This is reflected in the Onsager matrix, after expanding the $L_{ij}$ to first order in $dE$:
\begin{eqnarray}
L_{UU}&=&E_{0}^{2}L_{NN} \nonumber \\
L_{UN}&=&L_{NU}=E_{0}L_{NN} \nonumber \\
L_{NN}&=& \frac{\sigma \rho (E_{0}^{2}-m^{2}c^{4})e^{-E_{0}/kT}dE}{4km^{2}c^{3}K_2\left(mc^2/kT\right)}.
\end{eqnarray}
The determinant of the Onsager matrix is identically zero, as both rows are linear dependent. This opens up the possiblity, in principle, to use the system as a thermodynamic engine attaining Carnot or Curzon-Ahlborn efficiencies \cite{chris1,chris2}.

\section{Pair production}
One important aspect of relativistic gases, which we have not
considered so far, is the production and annihilation of
particle-antiparticle pairs \cite{HaWe}. This must be included in a
consistent treatment of relativistic effusion by considering the
corresponding exchange of antiparticles between the two
compartments.\newline 
In our framework, in which both reservoirs independently are (and remain) in equilibrium, this implies the existence of a second ideal gas in each reservoir consisting of anti-particles. Because of the production and annihilation, both particles and anti-particles will be created and destroyed. This process can be represented by a chemical reaction of the following form:
\begin{equation}
x + \bar{x} \rightleftharpoons 0,
\end{equation}
($x$ represents a particle, $\bar{x}$ the corresponding anti-particle). Equilibrium is achieved when the two chemical potentials, $\mu$ and $\bar{\mu}$, satisfy:
\begin{equation}\label{eq:chemeq}
\mu+\bar{\mu}=0.
\end{equation}
For a given temperature this relation determines the density $\bar{\rho}$ of the anti-particles. Using Eq.~(\ref{eq:chempot}) gives:
\begin{equation}\label{eq:densityeq}
\bar{\rho}=\left[4\pi \left(\frac{mc}{h}\right)^{3}\frac{K_{2}(mc^{2}/kT)}{mc^{2}/kT}\right]^{2}\frac{1}{\rho}.
\end{equation}
The same result can be obtained by a direct calculation of the grand canonical partition function of the two ideal gases, taking into account that for pair production and annihilation the difference between the
number of particles and antiparticles is conserved \cite{Huang}.\newline Note that fluctuations of (anti-)particles number and energy around their equilibrium values can be neglected, as both reservoirs are assumed to be infinitely large. \newline
The exchange of antiparticles through the opening between volumes A
and B will contribute to the entropy change $\Delta S$. Since the
chemical potential of the antiparticles is just the negative of the
chemical potential of the particles we find instead of
(\ref{eq:entropy}) 
\begin{eqnarray}
  \Delta S &=&\force_{U}\Delta U +\force_{N}\Delta N +\force_{\bN}\Delta\bN \nonumber \\
                &=&\force_{U}\Delta U +\force_{N}(\Delta N-\Delta\bN),
\end{eqnarray}
where $\force_{U}$ and $\force_{N}(=-\force_{\bN})$ are given by
(\ref{eq:forces}) and (\ref{eq:tforces}) respectively. To account for
the fluctuations in $\Delta S$ due to the exchange of antiparticles the joint
probability distribution has to be augmented to $P(\Delta U,\Delta N, \Delta \bN)$, and the corresponding cumulant generating function is given by: 
\begin{equation}
\langle e^{-(\lambda_{U}\Delta U+\lambda_{N}\Delta
  N+\lambda_{\bN}\Delta \bN)}\rangle=
  e^{-tQ(\lambda_{U},\lambda_{N},\lambda_{\bN})}\; .
\end{equation}
The detailed fluctuation theorem acquires the form 
\begin{multline}
Q(\lambda_{U},\lambda_{N},\lambda_{\bN} )=\\
Q(\force_{U}/k-\lambda_{U},\force_{N}/k-\lambda_{N},\force_{\bN}/k-\lambda_{\bN}).
\label{DFTpair}
\end{multline}
The time evolution of $P(\Delta U,\Delta N, \Delta \bN)$ is now
governed by the master equation:
\begin{widetext}
\begin{align}\nonumber
  \frac{\partial}{\partial t}P_{t}(\Delta U,\Delta N, \Delta \bN)
   =\int_{E_{1}}^{E_{2}} &\left[ 
       T_{A\rightarrow B}(E)P_{t}(\Delta U-E,\Delta N-1,\Delta \bN)
      +\bar{T}_{A\rightarrow B}(E)P_{t}(\Delta U-E,\Delta N,\Delta
         \bN-1)\right . \\\nonumber
      &+T_{B\rightarrow A}(E)P_{t}(\Delta U+E,\Delta N+1,\Delta \bN)
      +\bar{T}_{B\rightarrow A}(E)P_{t}(\Delta U+E,\Delta N,\Delta \bN+1)\\
      &\left.-P_{t}(\Delta U,\Delta N,\Delta \bN)
         \left(T_{A \rightarrow B}(E)+\bar{T}_{A \rightarrow B}(E)
         +T_{B \rightarrow A}(E)+\bar{T}_{B \rightarrow A}(E)\right)\right]dE, 
\label{eq:masterpair}
\end{align}
\end{widetext}
where the rates $T_{i\rightarrow j}(E)$ and $\bar{T}_{i\rightarrow j}(E)$ are defined by (\ref{eq:trrel}) with $\rho_i$ replaced by $\brho_i$ in the latter case. Similar to Eq.~(\ref{eq:mu}) we now find:
\begin{multline}
Q(\lambda_{U},\lambda_{N},\lambda_{\bN})=
   I_{A}(0,0)-I_{A}(\lambda_{U},\lambda_{N})
  +\bar{I}_{A}(0,0)\\ -\bar{I}_{A}(\lambda_{U},\lambda_{\bN}) 
  +I_{B}(0,0)-I_{B}(-\lambda_{U},-\lambda_{N}) \\
  +\bar{I}_{B}(0,0)-\bar{I}_{B}(-\lambda_{U},-\lambda_{\bN})\; ,
\label{eq:mupair1}
\end{multline}
with $I_i(\lambda_U,\lambda_N)$ given by Eq.~(\ref{eq:resI}). For $\bar{I}_i$ the density is $\bar{\rho_{i}}$.
Using these expressions in (\ref{eq:mupair1}) it is again straightforward to verify (\ref{DFTpair}). 

The inclusion of pair production thus opens an additional channel for
entropy production. However, due to the relations Eqs.~(\ref{eq:chemeq}-\ref{eq:densityeq})
between the chemical potentials of particles and antiparticles and
their respective concentrations in equilibrium the fluctuation theorem
nevertheless remains valid.

In the above, we have considered only one kind of
particle-antiparticle pair. The extension to different kinds of pairs
is straightforward. The contribution of the photons produced in the 
annihilation processes to the entropy production may be included along
the lines of \cite{cleurenEPL2007}.

\appendix
\section{Thermodynamical properties of a relativistic ideal gas}\label{thermo}
The basic thermodynamic properties follow directly from the partition function $Z$, which for a gas of $N$ identical non-interacting particles is given by:
\begin{equation}
Z=\frac{1}{N!}\left[\frac{1}{h^{3}}\int e^{-H(p)/kT}d^{3}xd^{3}p\right]^{N}.
\end{equation}
The Hamiltonian of a free relativistic particle is given by the usual expression of the relativistic energy:
\begin{equation}\label{hamil}
H(p)=\sqrt{p^{2}c^{2}+m^{2}c^{4}}.
\end{equation}
If the gas is enclosed in a volume $V$, the expression for $Z$ becomes:
\begin{eqnarray}
Z&=&\frac{1}{N!h^{3N}}\left[4\pi V \int_{0}^{\infty}p^{2}e^{-\sqrt{p^{2}c^{2}+m^{2}c^{4}}/kT}dp\right]^{N} \nonumber \\
&=&\frac{1}{N!}\left[4\pi V \left(\frac{mc}{h}\right)^{3}\frac{K_{2}(mc^{2}/kT)}{mc^{2}/kT}\right]^{N}\label{eq:Z},
\end{eqnarray}
where $K_{2}$ is the second modified Bessel function of the second kind.\newline
Given $Z$, it is now standard procedure to obtain the thermodynamic quantities of interest. The chemical potential $\mu$ can be derived from the Helmholtz free energy $F$ ($\rho=N/V$): 
\begin{equation}\label{eq:chempot}
\mu =\left.\frac{\partial F}{\partial N}\right\vert_{T,V} =-kT\ln\left(\frac{4\pi (mc)^{3}}{\rho h^{3}}\frac{K_{2}(mc^{2}/kT)}{mc^{2}/kT}\right),
\end{equation}
with $F=-kT\ln Z$ given by (in the thermodynamic limit):
\begin{equation}
F=-NkT\bigg[1+\ln\left(\frac{4\pi V (mc)^{3}}{Nh^{3}}\frac{K_{2}(mc^{2}/kT)}{mc^{2}/kT}\right) \bigg]. \label{eq:F}
\end{equation}

\section{Transition rates}\label{transition}
The calculation of the transition rates as given in Eq.~(\ref{eq:tr}), is similar to that of the pressure in textbooks on kinetic theory. $T_{i \rightarrow j}(E)dEdt$ is the probability to observe a particle with energy (kinetic + rest energy) in the range $[E,E+dE]$, crossing the hole from $i$ to $j$ during a time interval $dt$. The corresponding momentum of the particle must be in the range $[p,p+dp]$ with:
\begin{equation}
p=\frac{1}{c}\sqrt{E^{2}-m^2c^{4}}\;\;\;\; \mbox{and}\;\;\;\; dp= \frac{EdE}{c\sqrt{E^{2}-m^2}c^{4}}.
\end{equation}
Since the gas is in equilibrium, its momentum distribution has the usual Boltzmann form:
\begin{equation}
\phi_{i}(\vec{p})=\frac{mc^{2}/kT_{i}}{4\pi(mc)^{3}K_{2}(mc^{2}/kT_{i})}e^{-H(p)/kT_{i}}
\end{equation}
with $H(p)$ given in Eq.~(\ref{hamil}).
As the particle (with velocity $v$) must be able to reach the hole during the time interval $dt$, it must be located inside the volume $\sigma v\cos \theta dt$ ($\theta$ is the usual spherical coordinate, taking the $z$-axis perpendicular to the common wall between the two reservoirs and pointing from $i$ to $j$). Since the particle is moving towards the hole we have $0 \leq \theta \leq \pi/2$ . Adding all contributions from the different directions $\theta$ and $\varphi$ leads to the following result:
\begin{multline}
T_{i \rightarrow j}(E)dEdt \\ \quad=\int_{\theta=0}^{\pi /2}\int_{\varphi=0}^{2\pi}p^2 \sin \theta dpd\theta d\varphi\; \rho_{i}\phi_{i}(\vec{p})\;\sigma v\cos \theta dt.
\end{multline}
Finally, replacing $v=pc/\sqrt{p^{2}+mc^{2}}$ and integrating yields the expression given in Eq.~(\ref{eq:tr}).

\end{document}